\documentclass[iop]{emulateapj}

\usepackage{amsmath,natbib,graphicx}
\usepackage{epsf}
\usepackage{epstopdf}
\usepackage{epsfig}
\usepackage{color}
\DeclareGraphicsExtensions{.jpg,.pdf,.png,.eps,.ps}
\usepackage{scrextend}
\usepackage{hyperref}

\shorttitle{}

\begin{document}

\title{SOFIA/HAWC+ detection of a gravitationally lensed starburst galaxy at $z$ = 1.03}

\author{Jingzhe Ma$^{1}$, Arianna Brown$^{1}$, Asantha Cooray$^{1}$, Hooshang Nayyeri$^{1}$, Hugo Messias$^{2,3}$, Nicholas Timmons$^{1}$, Johannes Staguhn$^{4,5}$, Pasquale Temi$^{6}$, C. Darren Dowell$^{7}$, Julie Wardlow $^{8}$, Dario Fadda$^{9}$, Attila Kovacs$^{10}$, Dominik Riechers$^{11}$, Ivan Oteo$^{12}$, Derek Wilson$^{1}$, and Ismael Perez-Fournon$^{13}$}

\altaffiltext{1}{Department of Physics \& Astronomy, University of California, Irvine, 92697 USA; \href{mailto:jingzhem@uci.edu}{jingzhem@uci.edu}}
\altaffiltext{2}{Joint ALMA Observatory, Alonso de C\'ordova 3107, Vitacura 763-0355, Santiago, Chile}
\altaffiltext{3}{European Southern Observatory, Alonso de C\'ordova 3107, Vitacura, Casilla 19001, 19 Santiago, Chile}
\altaffiltext{4}{NASA Goddard Space Flight Center, Greenbelt, MD 20771, USA}
\altaffiltext{5}{Department of Physics \& Astronomy, Johns Hopkins University, Baltimore, MD, 21218, USA}
\altaffiltext{6}{NASA Ames Research Center, MS 245-6, Moffett Field, CA 94035, USA}
\altaffiltext{7}{Jet Propulsion Laboratory, MC 169-327, Pasadena, CA 91109, USA}
\altaffiltext{8}{Centre for Extragalactic Astronomy, Department of Physics, Durham University, South Road, Durham, DH1 3LE, UK}
\altaffiltext{9}{SOFIA Science Center, Universities Space Research Association, Moffett Field, CA 94035, USA }
\altaffiltext{10}{Smithsonian Astrophysical Observatory, MS-78, 60 Garden St, Cambridge, MA 02138, USA}
\altaffiltext{11}{Cornell University, Space Sciences Building, Ithaca, NY 14853, USA}
\altaffiltext{12}{Institute for Astronomy, University of Edinburgh, Royal Observatory, Blackford Hill, Edinburgh EH9 3HJ, UK}
\altaffiltext{13}{Instituto de Astrofisica de Canarias, E-38200 La Laguna, Tenerife, Spain}


\begin{abstract}

We present the detection at 89 $\micron$ (observed frame) of the {\it Herschel}-selected gravitationally lensed starburst galaxy HATLASJ1429-0028 (also known as G15v2.19) in 15 minutes with the High-resolution Airborne Wideband Camera-plus (HAWC+) onboard the Stratospheric Observatory for Infrared Astronomy (SOFIA). The spectacular lensing system consists of an edge-on foreground disk galaxy at $z$ = 0.22 and a nearly complete Einstein ring of an intrinsic ultra-luminous infrared galaxy at $z$ = 1.03. Is this high IR luminosity powered by pure star formation (SF) or also an active galactic nucleus (AGN)? Previous nebular line diagnostics indicate that it is star-formation dominated. We perform a 27-band multi-wavelength spectral energy distribution modeling (SED) including the new SOFIA/HAWC+ data to constrain the fractional AGN contribution to the total IR luminosity. The AGN fraction in the IR turns out to be negligible. In addition, J1429-0028 serves as a testbed for comparing SED results from different models/templates and SED codes ({\sc magphys}, {\sc sed3fit}, and {\sc cigale}). We stress that star formation history is the dominant source of uncertainty in the derived stellar mass (as high as a factor of $\sim$ 10) even in the case of extensive photometric coverage. Furthermore, the detection of a source at $z$ $\sim$ 1 with SOFIA/HAWC+ demonstrates the potential of utilizing this facility for distant galaxy studies including the decomposition of SF/AGN components, which cannot be accomplished with other current facilities. 

\end{abstract}


\section{Introduction}\label{sec:intro}

High-redshift dusty star-forming galaxies (DSFGs) are characterized by their very high intrinsic infrared (IR) luminosities $L_{\rm IR}$ $\sim$ 10$^{12}$ -- 10$^{13}$ $L_{\Sun}$ and dust enshrouded intense star formation with star formation rates (SFRs) exceeding $\sim$ 100 -- 1000 $M_{\Sun}$yr$^{-1}$ (for a review see \citealt{Casey2014}). Large area far-IR (FIR) to sub-millimeter/millimeter surveys such as the {\it Herschel} Astrophysical Terahertz Large Area Survey (H-ATLAS; \citealt{Eales2010,Bussmann2013}), the {\it Herschel} Multi-Tiered Extragalactic Survey (HerMES;  \citealt{Oliver2012,Bussmann2015}), and the South Pole Telescope (SPT) Sub-millimeter Galaxy (SMG) survey \citep{Vieira2013,Strandet2016, Spilker2016} have discovered large samples of gravitationally lensed DSFGs whose properties are being studied in unprecedented detail thanks to the significant gain in both brightness and spatial resolution provided by gravitational lensing \citep{Bussmann2013, Calanog2014, Spilker2016}.

DSFGs are in a unique phase of galaxy formation and evolution and provide a laboratory for studying the co-evolution of galaxies and their super massive black holes (SMBHs; \citealt{Alexander2012}). X-ray is arguably the best indicator of active galactic nuclei (AGN). X-ray observations of DSFGs show that the majority have bolometric luminosities that are dominated by star-formation rather than AGN emission (e.g., \citealt{Alexander2005a, Wang2013}). However, these galaxies can be highly obscured by large columns of dust and gas therefore difficult to be detected in the X-ray (e.g., \citealt{Alexander2005b}). Radio data provide another route to distinguishing star-forming galaxies from AGN. Star-forming galaxies follow a tight FIR-to-radio correlation over five orders of magnitude in galaxy luminosity, while radio-loud AGN produce excess radio emission above this relation (e.g., \citealt{Yun2001}). Alternatively, emission line diagnostics such as optical nebular emission lines can serve as indirect AGN indicators. However, rest-frame optical emission lines are often not detectable in these highly dust-obscured systems. None of these indicators provides a quantitative AGN fraction that can be used to correct for AGN contamination in measured $L_{\rm IR}$, SFR, and stellar mass, which are fundamental parameters that determine the nature of DSFGs.

Galaxy emission in the mid-IR contains rich information about the underlying sources within the galaxy, including polycyclic aromatic hydrocarbon (PAH) features (e.g., \citealt{Pope2008, Riechers2014}), which trace star formation, and additional hot dust emission from around any AGN. Combined with photometric data covering UV, optical, NIR and FIR, multi-wavelength spectral energy distribution (SED) modeling techniques are powerful tools to decompose star formation and AGN activity, and to quantitatively constrain the AGN fraction, in addition to providing self-consistent constraints on stellar masses, SFRs, stellar ages, and dust properties etc. \citep{daCunha2008, Noll2009, Kirkpatrick2012, Ciesla2015,Malek2017,Leja2017}. 

This paper focuses on a strongly lensed DSFG (magnification factor $\mu$ $\sim$ 10), HATLASJ142935.3-002836 (J1429-0028 hereafter; H1429-0028 in \citealt{Messias2014}; G15v2.19 in \citealt{Calanog2014}), standing out from the GAMA-15 field of {\it Herschel}-ATLAS \citep{Eales2010} due to its extremely bright fluxes with $S_{\rm 160\micron}$ = 1.1 $\pm$ 0.1 Jy. This system consists of an edge-on foreground disk galaxy at $z$ = 0.218 and a nearly complete Einstein ring of an intrinsic ultra-luminous infrared galaxy at $z$ = 1.027. The brightness enables detections from the optical to radio wavelengths, construction of lens models, and de-blending the source/lens photometry \citep{Calanog2014,Messias2014}. \cite{Messias2014} presented a detailed analysis of the gas and dust properties, dynamical information, and derive galaxy parameters from SED modeling of J1429-0028. However, a few critical questions remain unaddressed: 1) Is there an energetically-important AGN in the system and what is the fractional contribution to the total IR luminosity, if any? 2) How did the galaxy build up the stellar mass, i.e., are we able to constrain the star formation history (SFH)? Given that we have obtained extensive photometric coverage from UV to radio, the best case among all DSFGs, J1429-0028 can serve as a laboratory to test systematic uncertainties in deriving physical properties via panchromatic SED modeling.

In this paper, we present new photometric observations from the High-resolution Airborne Wideband Camera-plus (HAWC+;\citealt{Dowell2013,Smith2014}) onboard the Stratospheric Observatory for Infrared Astronomy (SOFIA; \citealt{Temi2014}), which is the only current facility that covers the 50 -- 100 $\micron$ wavelength range, a crucial regime for disentangling SF/AGN activity in galaxies at redshifts $z$ $\sim$ 0.5 - 2. In combination with our existing multi-wavelength data, we will address the remaining questions about J1429-0028 in order to probe the nature of this DSFG in unprecedented detail. The paper is organized as follows. The SOFIA observations and data reduction are presented in Section \ref{sec:data}. We perform multi-wavelength SED modeling including AGN templates in Section \ref{sec:sed}. The results are discussed in Section \ref{sec:results} and the conclusions are summarized in Section \ref{sec:conclusion}. We adopt a \cite{Chabrier2003} initial mass function (IMF) throughout the paper.  We assume a $\Lambda$CDM cosmological model with $H_{\rm 0}$ = 70 km s$^{-1}$ Mpc$^{-1}$, $\Omega_{\rm M}$ = 0.3, and $\Omega_{\Lambda}$ = 0.7. The IR luminosity is integrated over rest-frame 8-1000 $\micron$.

\section{Observations} \label{sec:data}

\subsection{SOFIA/HAWC+}

\begin{figure*} 
\centering
{\includegraphics[width=14cm, height=7cm]{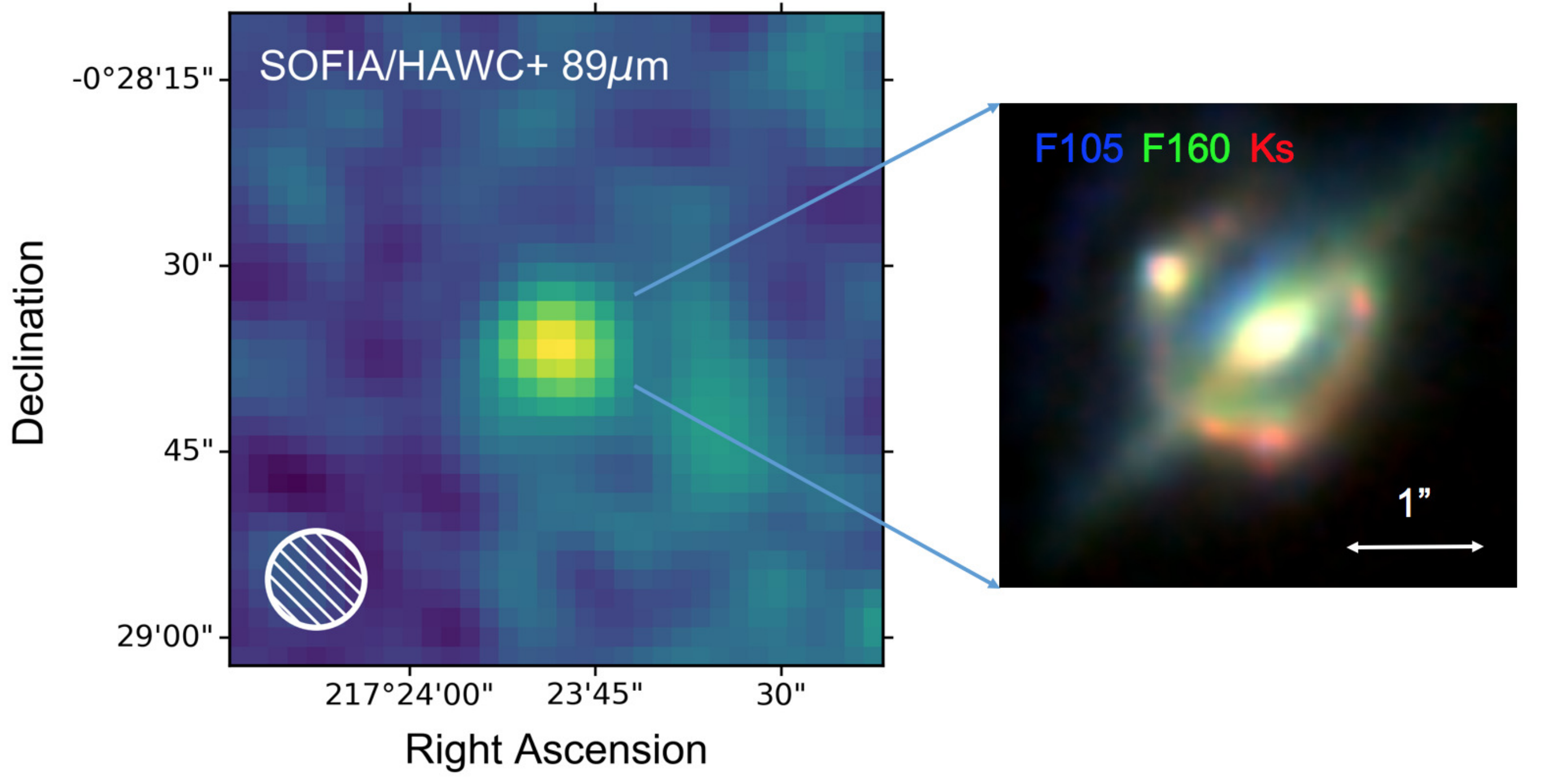}}  
\caption{{\it Left}: SOFIA/HAWC+ 89 $\micron$ detection of J1429-0028. The source is unresolved. The hatched circle shows the beam FWHM of 7.8$\arcsec$. {\it Right}: The 3-color image of the gravitationally lensed system using {\it HST} F105W (blue), F160W (green), and Keck Ks (red) imaging data \citep{Timmons2015}. }
\label{sofia}
\end{figure*}

J1429-0028 was observed with the HAWC+ instrument \citep{Dowell2013,Smith2014} onboard SOFIA on May 11th, 2017 under the Cycle 5 program PID05\_0087 (PI: Cooray). HAWC+ is a far-infrared imager and polarimeter with continuum bandpasses from 40 $\micron$ to 300 $\micron$. We obtained the observation in Band C at 89 $\micron$ (rest-frame 44 $\micron$) with a bandwidth of $\sim$ 17 $\micron$ in the Total-Intensity OTFMAP configuration (Figure \ref{sofia}). The image has a field of view of 4.2$\arcmin$ $\times$ 2.7$\arcmin$ and a resolution (FWHM) of 7.8$\arcsec$. The total effective on-source time is 939 s. The raw data was processed through the CRUSH pipeline v2.34-4 \citep{Kovacs2008}, and the final Level 3 data product was flux calibrated. The flux calibration error is about 10\%. The resulting map has an rms noise level of $\sim$ 50 mJy beam$^{-1}$. We extracted the photometry of the source using an 1-FWHM radius aperture and the resultant flux density is 748.4 $\pm$ 101.1 mJy. Using a lensing magnification factor of 9.7 $\pm$ 0.7 \citep{Messias2014}, the de-magnified flux density at 89 $\micron$ is 77.2 $\pm$ 11.8 mJy. The minor contribution to the observed flux density from the foreground lensing galaxy is estimated to be at most 3\%, which is within the quoted uncertainties.

\subsection{Multi-wavelength photometry}

Multi-wavelength photometry from rest-frame UV to radio has been obtained from SDSS ($u$, $g$, $r$, $i$, $z$), VIKING (Z, Y, J, H, K$_s$), {\it HST} (F110W), {\it Spitzer} (3.6, 4.5 $\micron$), {\it WISE} (3.4, 4.6, 12, 22 $\micron$), {\it Herschel} (100, 160, 250, 350, 500 $\micron$), IRAM-30 (1.2 mm), ALMA (1.28, 2.8 mm), and JVLA (7GHz) \citep{Messias2014}. The de-blending of the foreground lens and background source is discussed in detail in \cite{Messias2014}. We use the de-blended magnification-corrected photometry for the source to perform SED modeling.

\section{SED models}
\label{sec:sed}

\cite{Messias2014} performed SED fitting on J1429-0028 using the Multi-wavelength Analysis of Galaxy Physical Properties ({\sc magphys}; \citealt {daCunha2008}), which did not include AGN templates nor radio emission. In this work, we employ two SED fitting codes, {\sc sed3fit} \citep{Berta2013} and the Code Investigating GALaxy Emission ({\sc cigale} version v0.11; \citealt{Noll2009,Roehlly2014}), which include the additional components and rely on the energy balance technique, i.e., the energy of the absorbed starlight by dust is equal to the re-radiated energy by dust at infrared wavelengths. 

{\sc sed3fit} is a combination of the original {\sc magphys} code and the \cite{Fritz2006} AGN library (updated by \citealt{Feltre2012}) that is effectively fit to data in a simultaneous 3-component (stars, dust, AGN) model. Here we use the updated version of {\sc magphys} for high-$z$ IR models \cite{daCunha2015}, which extends the parameter space to include properties that are more likely applicable to high-redshift SMGs. {\sc cigale} includes up-to-date and customized star formation history models, various stellar population synthesis models and IR dust re-emission templates, AGN templates, and radio emission, to constrain the AGN contribution in the IR via an energy balance manner similar to {\sc sed3fit}. Both {\sc cigale} and {\sc sed3fit} fit galaxy SEDs using a Bayesian approach to generate the posterior probability distribution function (PDF) for each parameter of interest, marginalized over all other parameters. The output values of the analyzed parameters in {\sc cigale} are calculated as the weighted mean and standard deviation derived from the PDFs. {\sc sed3fit} outputs the 16\%, 50\% (median), 84\% values of the PDFs. The two codes are the most widely used panchromatic SED fitting codes, and whether or not they derive consistent results is of interest and may have impact on how we interpret the data. It has been suggested that multiple SED fitting approaches should be used to cross-check the results \citep{Hayward2015}. 

In the subsections below, we describe the various models employed in {\sc sed3fit} and {\sc cigale} respectively. Since there are more than one SFH, dust attenuation law, and dust emission models that one can easily change in {\sc cigale}, we will also test how different choices of models and parameters can affect the derived physical properties by varying one component at a time.

\subsection{Stellar component}

We adopt the \cite{Chabrier2003} initial mass function (IMF) and the stellar population synthesis models of \cite{Bruzual2003} (BC03) in both {\sc sed3fit} and {\sc cigale}. The original {\sc magphys} code, which \cite{Messias2014} used, implements the updated version of BC03 models (CB07), where the new prescription by \cite{Marigo2007} for the thermally pulsating asymptotic giant branch evolution of low- and intermediate-mass stars is incorporated. CB07 models have been tested to result in 50\% to 80\% lower stellar masses in HUDF galaxies than determined from the BC03 models \citep{Bruzual2007}.

\subsubsection{SED3FIT/MAGPHYS}
The original {\sc magphys} assumes an exponentially decreasing $\tau$ star formation history, where $\tau$ is the star formation e-folding time. However, \cite{Messias2014} adopted the models in \cite{Rowlands2014}, because the standard models have limitations to cover the physical parameter space for a DSFG. \cite{Messias2014} assumed both exponentially increasing and decreasing SFHs by distributing $\tau$ as a Gaussian between -1 and 1 Gyr$^{-1}$. Bursts of star formation are superimposed at random times on the continuous SFH with a 50\% probability of experiencing a burst in the last 2 Gyr. {\sc sed3fit} and the high-$z$ version of {\sc magphys} assume an underlying continuous star formation history with superimposed star formation bursts of random duration and amplitude. Each star formation burst can last between 30 to 300 Myr, and is set to occur at any random time in the previous 2 Gyr with a 75\% probability. The priors on metallicity are a uniform distribution from 0.2 to 2 times solar.

\subsubsection{CIGALE}
In {\sc cigale}, we first adopt the delayed-$\tau$ SFH model that rises in SFR up to a maximum, followed by an exponential decrease, which is motivated by high-redshift galaxies \citep{Lee2010}. Based on observations as well as galaxy evolution models for high-redshift galaxies \citep{Lee2010,daCunha2015}, the SFHs overall should be initially rising with time before declining, although the real SFHs can be much more complicated (e.g., \citealt{Behroozi2013, Pacifici2013, Simha2014, daCunha2015}). We also investigate the two-component $\tau$-SFH model, which contains a young and an old stellar populations (similar to the two components in {\sc sed3fit}/{\sc magphys} but in different forms), because real systems are likely to experience multiple episodes of star formation. There is evidence that the true stellar mass is likely better reproduced with the double SFHs \citep{Michalowski2014}, although the mass of the old stellar population is typically poorly constrained. Each stellar component is characterized by an exponentially declining SFH with two parameters: the stellar population age and the e-folding time. The two stellar populations are linked by the burst fraction (i.e., the mass fraction of the young stellar population). {\sc cigale} also allows self-defined SFH as input, but we do not attempt more complicated models. Possible values for metallicity are 0.0001, 0.0004, 0.004 (SMC), 0.008 (LMC), 0.02 (solar), and 0.05. 

\subsection{Dust attenuation}

\subsubsection{SED3FIT/MAGPHYS}
{\sc sed3fit}/{\sc magphys} uses the two-component dust attenuation model of \cite{Charlot2000} with each component described by a power-law. This model accounts for the fact that young stars in their birth clouds are more dust-attenuated than intermediate-age and old stars in the diffuse ISM. 

\subsubsection{CIGALE}
Three dust attenuation models can be used in {\sc cigale}: (1) a modified version of the \cite{Calzetti2000} attenuation law, (2) a single power-law model as defined in \cite{Charlot2000} with a UV bump added, (3) the two-component model of \cite{Charlot2000} as in {\sc sed3fit}/{\sc magphys}. {\sc cigale} enables both models (1) and (2) to allow differential attenuation in young and old stellar populations by using a reduction factor, which is defined as the ratio of the V-band attenuation of the old stellar population to the young population. We test all the dust attenuation laws and adopt the one that produces the best-fit SED.

\subsection{\it Dust emission}

\subsubsection{SED3FIT/MAGPHYS}
The dust emission in {\sc sed3fit}/{\sc magphys} consists of four components: (1) mid-IR continuum from host dust, (2) warm dust in thermal equilibrium, (3) cold dust in thermal equilibrium, and (4) PAH empirical templates. The cold and warm dust components in thermal equilibrium emit as modified black bodies with fixed emissivities. This dust model outputs the dust luminosity, dust mass, and dust temperatures for each component as well as a luminosity-weighted average dust temperature.

\subsubsection{CIGALE}

We first employ \cite{Dale2014} IR dust emission models (updated \citealt{Dale2002} models), which have been successfully applied to SMGs \citep{Ma2015,Ma2016}. The semi-empirical \cite{Dale2014} templates are parameterized by the power-law slope of the dust mass distribution over heating intensity $\alpha$, $dM/dU$ $\propto$ $U^{-\alpha}$. The dust emission templates are connected to the attenuated stellar population models by the dust luminosity $L_{\rm dust}$, which sets the basis of energy balance.

We also test other physically-motivated IR models. We use an updated version of the \cite{Draine2007} model that is implemented in {\sc cigale} \citep{Aniano2012, Ciesla2014}. The \cite{Draine2007} model is motivated by the physical nature of dust (composition, geometry, and size distribution), and parameterized by the PAH mass fraction, minimum and maximum intensities of the interstellar radiation field, and the relative dust mass fraction heated by the diffuse radiation and the photodissociation regions (PDRs). The normalization of the model to the photometric data yields the dust mass, $M_{\rm dust}$, in addition to $L_{\rm dust}$. 

\subsection{AGN emission}

The AGN emission in both SED codes is based on the AGN templates from \cite{Fritz2006}, which consist of two components. One is the isotropic emission of the central AGN and the other is an improved model of the emission from the dusty torus heated by the central engine. Part of the central emission is absorbed by the dusty torus and re-emitted at longer wavelengths. {\sc sed3fit} also includes the emission of the accretion disk in an updated library by \cite{Feltre2012}. The full library comprises 2376 models with each one computed at 10 different lines of sight angles. {\sc sed3fit} provides a set of 10 AGN templates spanning the whole range of colors covered by the full library. A few parameters are specified to characterize the dust torus, including inner/outer radius, radial and angular dust distributions, angular opening angle etc. Type 1 (unobscured), Type 2 (obscured), as well as intermediate-type AGN can be approximately described by those parameters. The most important parameter is the fractional contribution of AGN to the total IR luminosity ($L_{\rm IR, total}$ = $L_{\rm Starburst}$ + $L_{\rm AGN}$). We refer the reader to \cite{Berta2013} and \cite{Malek2017} for the detailed implementation and allowed parameter ranges in each code.

\subsection{Radio emission}

The SED fit is also extended to radio wavelengths based on the well-established FIR-to-radio correlation, $q$, for star-forming galaxies. A tight FIR-to-radio correlation has been observed for star-forming galaxies over five orders of magnitude in galaxy luminosity (e.g., \citealt{Yun2001}). Radio-loud AGN, on the contrary, exhibit elevated radio emission that would be at least a factor of $\sim$ 2 higher \citep{Rush1996,Moric2010}. This radio emission model can be used to test whether or not the observed radio photometric data is consistent with the FIR-to-radio correlation.

{\sc sed3fit}/high-$z$ {\sc magphys} computes the radio emission as the sum of a thermal (free-free emission) and a non-thermal component with fixed power-law slopes. A Gaussian prior distribution is assumed for the coefficient, $q$, with the mean at 2.34 (local value; \citealt{Yun2001}) and 1$\sigma$ value of 0.25 (e.g., \citealt{Ivison2010}). In {\sc cigale}, a non-thermal (synchrotron emission from galaxies) component is added and the thermal emission is handled by the nebular module. The default coefficient of the FIR/radio correlation is 2.58 and the spectral index of the synchrotron power-law emission is 0.8 (same as in {\sc sed3fit}). We test different prior values of 2.2 - 2.6 for $q$ as it does not necessarily follow the local relation and may evolve with redshift \citep{Delhaize2017}.

\begin{figure*}
\centering
{\includegraphics[width=16cm, height=12cm]{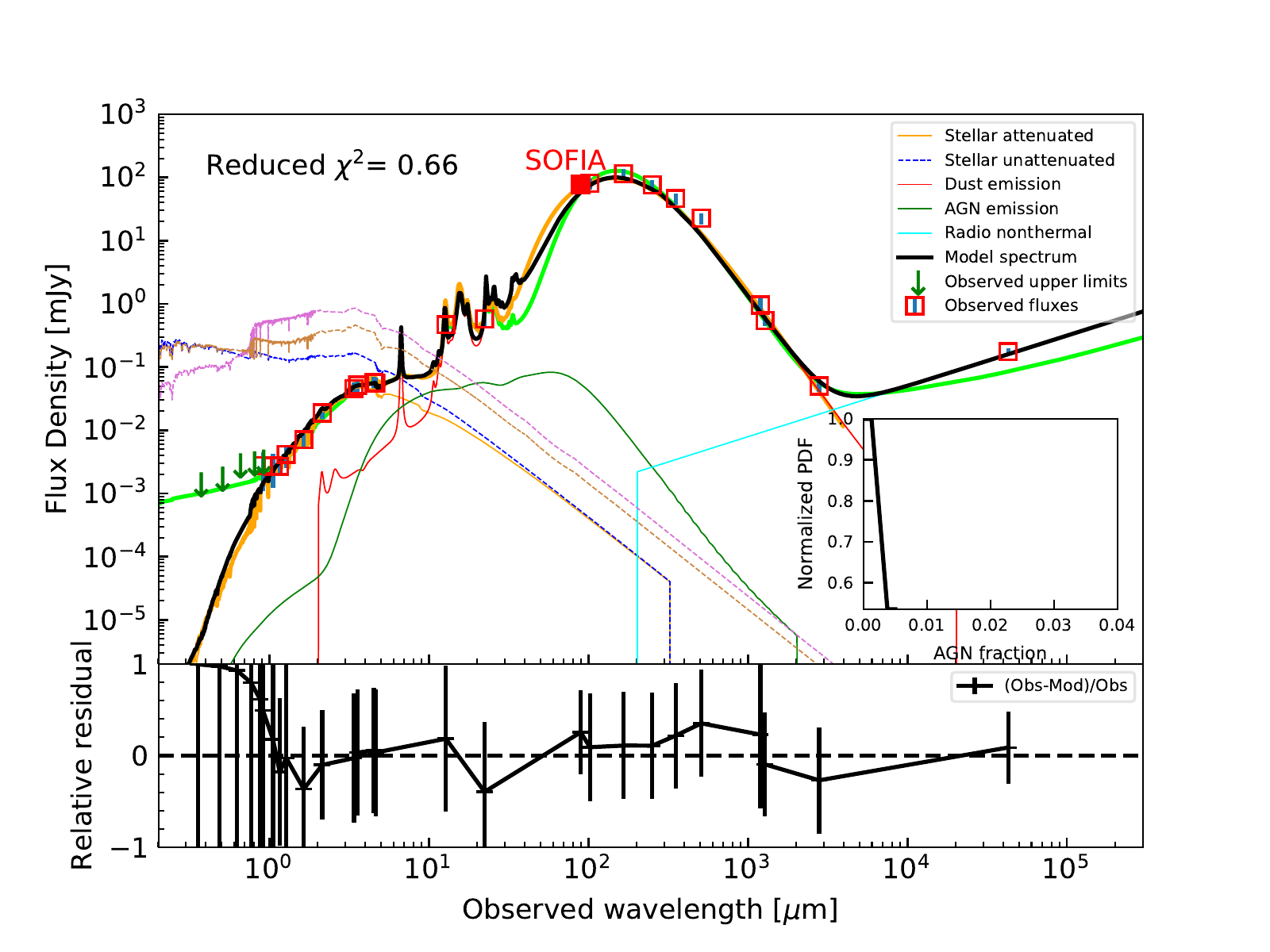}}  
\caption{The best-fit SED from {\sc cigale} (black; Model E) compared to the best-fits from {\sc sed3fit} (green; Model C) and {\sc magphys} (orange; Model A). Different {\sc cigale} SED components are color-coded. The unattenuated stellar emission from {\sc sed3fit} (purple) and {\sc magphys} (brown) is also displayed in comparison with the one from {\sc cigale} (blue). The data points from left to right are SDSS $u$$g$$r$$i$ and VIKING Z 3 $\sigma$ upper limits, SDSS $z$, VIKING Y, {\it HST}/F110W, VIKING J, H, K$_s$, WISE1, {\it Spitzer}/IRAC 3.6, 4.5 $\micron$, WISE2, WISE3, WISE4, SOFIA/HAWC+ 89 $\micron$ (filled red square), {\it Herschel}/PACS 100, 160 $\micron$, {\it Herschel}/SPIRE 250, 350, 500 $\micron$, IRAM-30 1.2 mm , ALMA 1.28, 2.8 mm, and JVLA 7 GHz.  The inset shows the normalized probability distribution function of the AGN fraction in the total IR luminosity.  The bottom panel shows the relative residual fluxes of the {\sc cigale} SED fit. }
\label{sed}
\end{figure*}

\begin{table}
\caption{Input photometry for SED fitting with MAGPHYS, SED3FIT and CIGALE. } 
\label{results}
\centering
\begin{tabular}{lcc}
\hline\hline
 Survey/facility   & Filters                              & Flux density (mJy)\\
\hline
SDSS       &      $u$           &   $<$ 1.36 $\times$ 10$^{-3}$        \\
                 &      $g$           &   $<$ 1.67 $\times$ 10$^{-3}$         \\
                 &      $r$           &    $<$ 2.72 $\times$ 10$^{-3}$        \\
                 &      $i$           &    $<$ 2.93 $\times$ 10$^{-3}$        \\
                 &      $z$           &   (2.77 $\pm$ 1.69) $\times$ 10$^{-3}$        \\     
VIKING     &       Z              &  $<$ 3.12 $\times$ 10$^{-3}$          \\
                 &       Y              &    (2.72 $\pm$ 1.51) $\times$ 10$^{-3}$       \\
                &       J             &     (4.15 $\pm$ 1.62) $\times$ 10$^{-3}$       \\  
                &       H              &  (7.08 $\pm$ 1.60) $\times$ 10$^{-3}$          \\  
                &       K$_s$              &    (1.88 $\pm$ 0.37) $\times$ 10$^{-2}$        \\    
HST          &      F110W       &  (2.69 $\pm$ 0.72) $\times$ 10$^{-3}$           \\
{\it Spitzer}   &  3.6$\mu$m   & (5.17 $\pm$ 1.18) $\times$ 10$^{-2}$  \\
                     & 4.5$\mu$m  & (5.69 $\pm$ 1.30) $\times$ 10$^{-2}$  \\
WISE         &   3.4$\mu$m     &   (4.57 $\pm$ 1.07) $\times$ 10$^{-2}$ \\
                   &   4.6$\mu$m     &  (5.53 $\pm$ 1.28) $\times$ 10$^{-2}$ \\  
                   &   12$\mu$m     & (4.68 $\pm$ 1.24) $\times$ 10$^{-1}$  \\    
                   &   22$\mu$m     &  (5.80 $\pm$ 1.48) $\times$ 10$^{-1}$\\       
SOFIA      &       89$\mu$m    &  77.2 $\pm$ 11.8   \\           
{\it Herschel} & 100$\mu$m     &  81.0 $\pm$ 15.9 \\       
                      & 160$\mu$m     & 114.8 $\pm$ 22.3 \\  
                      & 250$\mu$m     &76.7 $\pm$ 14.7  \\   
                      & 350$\mu$m     & 46.3 $\pm$ 8.9 \\    
                      & 500$\mu$m     & 22.6 $\pm$ 4.4 \\      
IRAM-30        & 1.2mm            & 0.95 $\pm$ 0.26\\
ALMA             & 1.28mm  & (5.43 $\pm$ 1.02) $\times$ 10$^{-1}$ \\
                      & 2.8mm    & (5.03 $\pm$ 0.97) $\times$ 10$^{-2}$ \\
JVLA              &  7GHz    &(1.75 $\pm$ 0.23) $\times$ 10$^{-1}$ \\                                                  
  \hline
\end{tabular}
\tablecomments{We note that the background source flux densities listed in Table 4 of \cite{Messias2014} are the best-fit model flux densities derived from {\sc magphys} rather than the input photometry.}
\end{table}

\begin{table*}
\caption{Physical properties derived from MAGPHYS, SED3FIT and CIGALE } 
\label{results}
\centering
\begin{tabular}{lcccccc}
\hline\hline
 Models \& Properties                            &  A. MAGPHYS*                              & B.  high-$z$ MAGPHYS                & C. SED3FIT  			       & D. CIGALE                     & E. CIGALE                      & F. CIGALE\\
 \hline
 Stellar population                                  & CB07                                       & BC03                            	            & BC03   						& BC03				& BC03		    & BC03\\
 SFH              				                   &  modified $\tau$ SFH  	                     & delayed-$\tau$ SFH    	        & delayed-$\tau$ SFH              &  delayed-$\tau$ SFH & double-$\tau$ SFH     & double-$\tau$ SFH \\   
                                                             &  + random bursts                       & + random bursts  	  		&    + random bursts                 &                                   & ($f_{\rm burst}$ = 0-1)&  ($f_{\rm burst}$ $\leq$ 0.1)  \\[0.07cm]   
Dust attenuation                                  &    Charlot+2000                  & Charlot+2000                       & Charlot+2000                   &   power-law   & power-law  &              power-law\\
 Dust emission                                      &  Rowlands+2014              & daCunha+2015                  &   daCunha+2015               & Draine+2014     & Draine+2014      & Draine2014\\     
 AGN emission                                       & ---     					    & ---						& Feltre+2012    			& Fritz+2006 	 &Fritz+2006		&Fritz+2006\\     [0.07cm]        
\hline
  SFR ($M_\sun$\,yr$^{-1}$)                &      ---  				              & ---    						& 161{\small $^{+112}_{-61}$}  & 714 $\pm$ 75             & 582 $\pm$ 29             & 516 $\pm$ 30       \\[0.07cm]
SFR$_{\rm 10Myr}$($M_\sun$\,yr$^{-1}$)&  ---                                        & ---           					& 163{\small $^{+111}_{-61}$}  & 662 $\pm$ 56   	   & 582 $\pm$ 29	         & 516 $\pm$ 30\\[0.07cm]
SFR$_{\rm 100Myr}$($M_\sun$\,yr$^{-1}$)&394{\small $^{+91}_{-88}$} &225{\small $^{+108}_{-107}$}     & 171{\small $^{+113}_{-67}$}  & 361 $\pm$ 36            & 296 $\pm$ 21		&266 $\pm$ 13 \\[0.07cm]
  $M_*$ (10$^{10}$ M$_\sun$)  	  & 13.2{\small $^{+6.3}_{-4.1}$}    &  44{\small $^{+26}_{-18}$}         & 83{\small $^{+43}_{-29}$}      & 2.1 $\pm$ 0.5             & 3.4 $\pm$ 1.1             &  15.1 $\pm$ 1.2 \\[0.07cm]
  sSFR (Gyr$^{-1}$)                  	           &   3.0 $\pm$ 1.6                          &   0.51 $\pm$ 0.39                	&  0.21 $\pm$ 0.17                   & 17.2 $\pm$ 4.4            &   8.7 $\pm$ 2.9                  & 1.8 $\pm$ 0.2  \\[0.07cm]
  age$_{\rm M}$ (Myr)                                            & 258{\small $^{+294}_{-111}$}   & 948{\small $^{+879}_{-544}$}    &1496{\small $^{+617}_{-552}$}&   $<$ 160  		    & $<$ 600 		          & 1991 $\pm$ 314         \\[0.07cm]
  $A_{\rm V}$                                        & ---                                              & 5.6{\small $^{+0.4}_{-0.5}$}       & 7.1{\small $^{+0.4}_{-0.9}$} 	& 4.3 $\pm$ 0.2            & 4.2 $\pm$ 0.2               & 4.1 $\pm$ 0.2   \\[0.07cm]
  $L_{\rm dust}$ (10$^{12}$ $L_\sun$) &4.27{\small $^{+0.63}_{-0.55}$} & 4.17{\small $^{+0.51}_{-0.62}$} &4.57{\small $^{+0.22}_{-0.40}$}&4.59 $\pm$ 0.23       & 4.41 $\pm$ 0.22           & 4.13 $\pm$ 0.21          \\[0.07cm]
  $M_{\rm dust}$  (10$^{8}$ $M_\sun$)&3.9{\small $^{+0.6}_{-0.6}$}       & 2.9{\small $^{+0.7}_{-0.4}$}       & 2.6{\small $^{+0.3}_{-0.4}$}     & 6.4 $\pm$ 0.4          &   6.3 $\pm$ 0.5             &   6.2  $\pm$ 0.6     \\[0.07cm]
  $T_{\rm dust}$ (K)                               &---					      &  40.7{\small $^{+2.8}_{-2.7}$}   &  40.7{\small $^{+1.3}_{-0.6}$}   & --- 		             &     ---   			  &   ---   \\[0.07cm]
  $f^{\rm AGN}_{\rm IR}$   (\%)              &   0                                             &   0        	        				 &0.16{\small $^{+0.06}_{-0.06}$}& $<$ 0.90                  &   $<$ 0.72   		  &   $<$ 0.75   \\[0.07cm]
  $\chi^2_{\rm red}$                               & 0.50                                           & 0.43						 &     0.56   				  &  0.63                         &   0.66     			  & 0.87   \\[0.07cm]
  \hline
\end{tabular}
\tablecomments{\noindent SFR is the current star formation rate. SFR$_{\rm 10Myr}$ is the SFR averaged over the last 10 Myr. SFR$_{\rm 100Myr}$ is the SFR averaged over the last 100 Myr. age$_{\rm M}$ is the mass-weighted stellar population age. sSFR is based on SFR$_{\rm 100Myr}$. $T_{\rm dust}$ is the weighted dust temperature of the birth cloud and ISM dust temperatures. $f^{\rm AGN}_{\rm IR}$ from CIGALE is given at 3 $\sigma$. $\chi^2_{\rm red}$ is the reduced $\chi^2$ from the SED fitting. *Previous SED modeling results from \cite{Messias2014} are listed here for reference. \cite{Messias2014} adopted the models in \cite{Rowlands2014}, because the standard models have limitations to cover the physical parameter space for J1429-0028. }
\end{table*}

\section{Results and Discussion}
\label{sec:results}

Table \ref{results} lists the properties derived from the PDF analysis of {\sc sed3fit} and {\sc cigale}, including the instantaneous SFR, SFRs averaged over the last 10 Myr and 100 Myr, $M_*$, stellar population age (mass-weighted age), V-band extinction, IR luminosity, dust mass, dust temperature, and the AGN fraction in the IR luminosity. {\sc sed3fit} also calculates the luminosity-averaged dust temperature as an output. Using the SFR and $M_*$, we also derive the specific SFR (sSFR) which is a critical diagnostic for understanding the star formation mode of this galaxy.

We have checked the PDFs to make sure that the way {\sc sed3fit} and {\sc cigale} calculate the average values of parameters, either median ({\sc sed3fit}) or weighted mean ({\sc cigale}), does not make much difference (i.e., within 1 $\sigma$ uncertainty).

\subsection{Star formation history}

A single stellar population with a delayed-$\tau$ SFH model from {\sc cigale} (Model D) yields the highest SFR and the lowest $M_*$ among the model configurations. The stellar population age is poorly constrained and the upper limit is consistent with the timescale of a recent starburst during the SMG phase (e.g., \citealt{Tacconi2006,Ivison2011}). 

The double SFH model from {\sc cigale} (Model E) produces a higher stellar mass and a lower current SFR. The initial input burst fraction is set as a free parameter, i.e., 0-1. The SEDs favor an almost pure young burst (burst fraction $\sim$ 0.8) for $\sim$100 Myr, suggesting that the stellar mass in the entire system is built up rapidly in the last 100 Myr. The prodigious star formation is most likely triggered by major mergers, based on the source-plane morphology and dynamical mass estimates \citep{Messias2014}. A high burst fraction further supports this scenario. However, this burst fraction derived from SED modeling is significantly higher than those from merger simulations ($<$ 50\%; \citealt{Hopkins2013}). \cite{Smith2015} use merger simulations to test how well {\sc magphys} is able to recover SFHs, where the true SFH is known. They fail to obtain a good estimate of the SFH for the merger simulations despite being able to get a reasonable fit to the synthetic photometry. Likewise, we test {\sc cigale} using merger simulations to check whether the burst fraction can be reasonably recovered. The comparison suggests that {\sc cigale} tends to overestimate the burst fraction likely due to the similar effect of the ``outshining'' problem of the young stellar populations over older stellar populations (e.g. \citealt{Reddy2012}). If we restrict the burst fraction to be $\leq$ 0.1 (Model F; typical burst fraction suggested by merger simulations \citealt{Hopkins2013}), the resultant stellar mass increases by a factor of $\sim$ 5 without much change in the SFR or the quality of the fit, and gets into agreement with the results in \cite{Messias2014}.

The {\sc sed3fit} model (Model C) with delayed-$\tau$ SFH and random bursts yield the highest stellar mass and lowest SFR. The blue end of the best-fit SED from SED3FIT shown in Figure \ref{sed} is much higher than {\sc magphys} and {\sc cigale} best-fit SEDs due to the unconstraining upper limits, and is dominated by AGN emission, although there is no evidence of a Type 1 AGN in the system \citep{Timmons2015}. This significantly affects the stellar SED and the mass-to-light ratio, resulting in a much higher stellar mass and lower SFR.

\cite{Wardlow2011} test the effect of SFHs on stellar masses and ages for SMGs with 17-band photometry. They cannot reliably distinguish the different SFHs and estimate that this results in an additional factor of $\sim$ 5 uncertainty in the mass-to-light ratios and therefore the derived stellar masses. Even in the case of extensive photometric coverage such as J1429-0028, we cannot draw definitive conclusions about the star formation history (stellar ages, burst fractions) solely based on the SED modeling.

\subsection{Dusty system}

All the SED fits indicate high dust extinction, $A_{\rm V}$ $\sim$ 4-7 and ultra-luminous dust emission $>$ 4 $\times$ 10$^{12}$ $L_\sun$. The dust luminosity is the best constrained parameter with least uncertainty; the dust luminosities derived from different models and SED codes are all consistent with each other. The best-fit FIR SEDs from {\sc magphys}, {\sc sed3fit}, and {\sc cigale} are similar while the MIR slopes are different due to lack of data. The derived dust masses from different models and SED codes are in the range of $\sim$ 3-6 $\times$ 10$^{8}$ $M_\sun$. The {\sc cigale} FIR dust emission is based on the updated \cite{Draine2007} model, and we adopt in {\sc sed3fit} the high-$z$ version of the dust emission model from {\sc magphys} \citep{daCunha2015}, because the original (low-$z$) library does not properly constrain the physical properties for this kind of galaxy. The average dust temperature is about 40 K, which is consistent with that of star-formation dominated DSFGs.

\subsection{Negligible AGN contribution}

\begin{figure}
\centering
{\includegraphics[width=9cm, height=7cm]{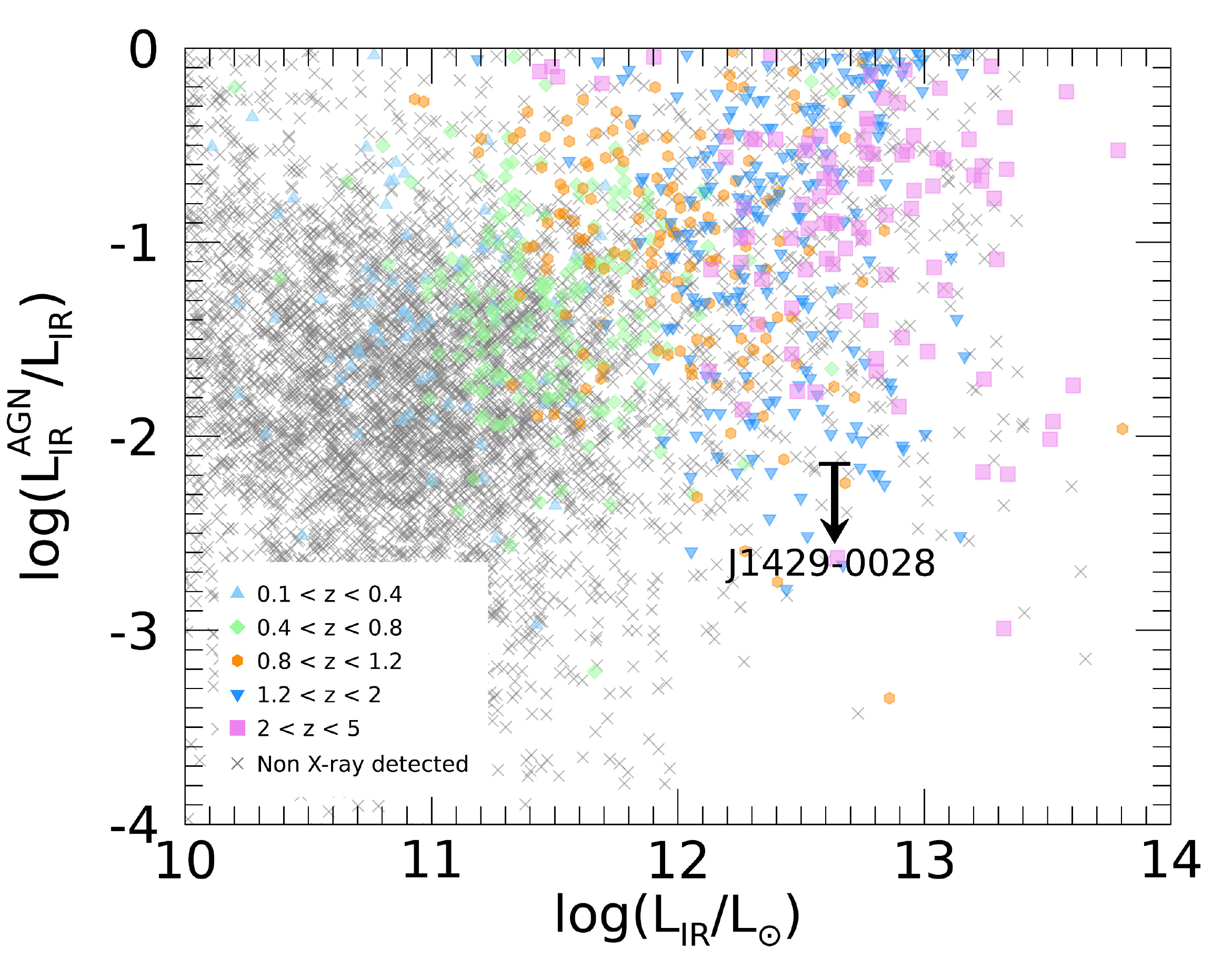}}  
\caption{AGN fraction $f^{\rm AGN}_{\rm IR}$ versus $L_{\rm IR}$. The colored symbols are the X-ray detected AGN in different redshift bins by \cite{Brown2018}, which represents the largest AGN sample with multi-wavelength SED fitting analysis. The grey X's are the X-ray non-detections. There is a general trend of higher AGN fraction with increasing $L_{\rm IR}$. The downward arrow denotes the $f^{\rm AGN}_{\rm IR}$ 3 $\sigma$ upper limit of J1429-0028.   }
\label{fagn}
\end{figure}

Previous nebular line diagnostics show no indication of an unobscured Type 1 AGN in J1429-0028 \citep{Timmons2015}. However, an obscured Type 2 AGN could potentially contribute to the observed very high $L_{\rm IR}$. We set the input AGN type as a free parameter, i.e., Type 1, Type 2 and intermediate type AGN. Regardless of the different combinations of component models and SED codes, the AGN contribution to the IR luminosity is negligible (3 $\sigma$ upper limit $f^{\rm AGN}_{\rm IR}$ $<$ 0.72\%). We note, however, that the best-fit SED in the blue end from {\sc sed3fit} (green) is significantly higher than the ones from {\sc magphys} (gold) and {\sc cigale} (black) due to an unexpected Type 1 AGN component dominating the UV spectrum and also affecting the optical-NIR part. This is simply caused by the unconstraining upper limits in the blue end. We cannot rely on the SED fitting alone to infer AGN activity in this case (more discussion in Section \ref{systematic}).

The JVLA radio data at 7GHz is also consistent with the FIR-to-radio correlation for star-forming galaxies. Radio-loud AGN would push the radio emission at least a factor of 2 higher.  We note that in this particular galaxy the new SOFIA data at 89 $\micron$ does not make much difference given the presence of PACS 100 $\micron$. A 53 $\micron$ band is preferred for constraining the slope of the MIR SED. For sources that do not have PACS data, the 89 $\micron$ band can be critical.

In Figure \ref{fagn} we compare the AGN fraction and IR luminosity of J1429-0028 with those of {\it Chandra} X-ray selected AGN in the Bo$\ddot{o}$tes legacy field \citep{Jannuzi1999,Murray2005} with MIR and FIR counterparts detected by {\it Spitzer} \citep{Ashby2009} and {\it Herschel} \citep{Oliver2012}, which represents the largest AGN sample with multi-wavelength SED fitting analysis \citep{Brown2018}. Figure \ref{fagn} shows the general trend of higher AGN fraction with increasing IR luminosity. The colored symbols denote the X-ray detected AGN in different redshift bins. The gray X's are the X-ray non-detections in the survey. J1429-0028 has an intrinsic IR luminosity that is higher than most of the sources at similar redshifts, comparable to the average $L_{\rm IR}$ at higher redshifts. The average (median) AGN fraction in the X-ray selected sample in the redshift bin 0.8 $<$ $z$ $<$ 1.2 is $\sim$10\%, while the 3$\sigma$ upper limit of J1429-0028 is significantly below the majority of the X-ray detected AGN. \newline

\subsection{Systematic uncertainties in SED modeling}
\label{systematic}

Given that the AGN fraction is negligible in the IR, we now compare the other critical physical properties with and without the inclusion of the AGN component. Previous best effort on the SED modeling of this source \citep{Messias2014} was based on the original {\sc magphys}, which employs the CB07 stellar population synthesis models rather than the BC03 models (used in the new high-$z$ version of {\sc magphys}). The inferred stellar masses from the BC03 models are about 1.25 - 1.5 times higher than from the CB07 models \citep{Bruzual2007}. The high-$z$ version of {\sc magphys} (Model B) results in a factor of $\sim$ 3 higher stellar mass for J1429-0028. The additional factor of $\sim$ 2 may be due to a combination of different choices of SFHs and IR dust emission models. We further compare the results from the high-$z$ {\sc magphys} (Model B) and {\sc sed3fit} (Model C) for which the only difference is the inclusion of the AGN component. The physical parameters are consistent with each other within 1$\sigma$ uncertainties. However, we stress that one should use multiple AGN indicators otherwise an unexpected AGN component can appear in the SED in the case of loose photometric data constraints in the blue end, which affects the derived stellar mass, SFR etc. Deriving reliable stellar masses from SED fitting for DSFGs has been extremely challenging. We have demonstrated that even in the case of extensive photometric coverage from UV to radio for a single galaxy, the systematic uncertainties can be as high as a factor of $\sim$10. Therefore, one should explicitly state the models involved in SED modeling especially the SFHs and convert to similar assumptions if possible whenever we compare stellar mass measurements.

\section{Summary and Conclusion}
\label{sec:conclusion}

We have presented the SOFIA/HAWC+ 89$\micron$ detection of the gravitationally lensed DSFG at $z$ = 1.027.
We conduct a detailed analysis of multi-wavelength SED modeling from rest-frame UV to radio, including the new SOFIA/HAWC+ data, and test how different combinations of component models or parameters and SED codes affect the derived physical properties. A significant AGN contribution to the IR luminosity has been ruled out at a high confidence level regardless of model choices. A significant fraction of the total stellar mass $\sim$ 10$^{11}$ $M_\sun$ in the galaxy is built up in rapid starburst in the last $\sim$ 100 Myr with an instantaneous SFR of $>$ 500 $M_\sun$\,yr$^{-1}$, likely triggered by a major merger. The inferred specific SFR places J1429-0028 above the star-forming main sequence at $z$ $\sim$ 1, further indicating the strong starbursting nature of the galaxy. 

We stress that even in the case of extensive photometric coverage, the uncertainty in stellar mass of a DSFG can be as high as a factor of $\sim$10 and is dominated by the assumed SFH. Therefore it is of crucial importance to state explicitly the model assumptions associated with the derived stellar masses. One should also combine other AGN indicators with SED fitting including an AGN component to infer the AGN activity.

SOFIA/HAWC+ is currently the only facility that covers the wavelength regime 50 -- 100 $\micron$ and such observations become critical for targets that do not have coverage in that spectral region. Although this particular galaxy does not contain an energetically-important AGN, this paper demonstrates the potential of using SOFIA/HAWC+ to constrain the AGN fraction in {\it Herschel}-selected galaxies due to lack of MIR data.

\section*{acknowledgments}

We thank the anonymous referee for the very constructive comments that have greatly improved the manuscript. We thank Christopher Hayward and Desika Narayanan for helpful discussions. This work is based on observations made with the NASA/DLR Stratospheric Observatory for Infrared Astronomy (SOFIA). SOFIA is jointly operated by the Universities Space Research Association, Inc. (USRA), under NASA contract NAS2-97001, and the Deutsches SOFIA Institut (DSI) under DLR contract 50 OK 0901 to the University of Stuttgart. Financial support for this work was provided by NASA through project PID05\_0087. J.L.W. acknowledges the support of an STFC Ernest Rutherford Fellowship under grant number ST/P004784/1. D.R. acknowledges support from the National Science Foundation under grant number AST-1614213. This research has made use of NASA's Astrophysics Data System.

\end{document}